\documentclass[prd,twocolumn,twoside,preprintnumbers,superscriptaddress,nofootinbib]{revtex4}
\usepackage{amsmath,slashed,dsfont}
\usepackage{graphicx,graphics,color}
\usepackage{dcolumn}
\usepackage[hyperfootnotes=false]{hyperref}
\usepackage{xspace}

\newcommand{\Fpi}{F_\pi}
\newcommand{\mpi}{M_{\pi}}

\newcommand{\MeV}{\,\text{MeV}}
\newcommand{\GeV}{\,\text{GeV}}
\newcommand{\mk}{M_K}
\newcommand{\diff}{\text{d}}
\newcommand{\mN}{m_N}

\newcommand{\beq}{\begin{equation}}
\newcommand{\eeq}{\end{equation}}
\newcommand{\unity}{\mathds{1}}
\newcommand{\fm}{\,\text{fm}}

\renewcommand{\Im}{\text{Im}\,}

\begin{document}

\preprint{INT-PUB-18-057}
\title{Nucleon matrix elements of the antisymmetric quark tensor}

\author{Martin Hoferichter}
\affiliation{Institute for Nuclear Theory, University of Washington, Seattle, WA 98195-1550, USA}
\author{Bastian Kubis}
\affiliation{Helmholtz--Institut f\"ur Strahlen- und Kernphysik (Theorie) and\\
   Bethe Center for Theoretical Physics, Universit\"at Bonn, 53115 Bonn, Germany}
\author{Jacobo Ruiz de Elvira}
\affiliation{Albert Einstein Center for Fundamental Physics, Institute for Theoretical Physics,
University of Bern, Sidlerstrasse 5, 3012 Bern, Switzerland}   
\author{Peter Stoffer}
\affiliation{Department of Physics, University of California at San Diego, La Jolla, CA 92093, USA}

\begin{abstract}
If physics beyond the Standard Model enters well above the electroweak scale, its low-energy effects are described by 
Standard Model Effective Field Theory. Already at dimension six many operators involve the antisymmetric quark tensor $\bar q \sigma^{\mu\nu} q$, whose matrix elements
are difficult to constrain from experiment, Ward identities, or low-energy theorems, in contrast to the corresponding vector and axial-vector or even scalar and pseudoscalar currents. 
However, with normalizations determined from lattice QCD, analyticity and unitarity  often allow one to predict the momentum dependence in a large kinematic range. Starting from 
recent results in the meson sector, we extend this method to the nucleon case and, in combination with pole dominance, provide a comprehensive assessment of the current status of
the nucleon form factors of the quark tensor.
\end{abstract}

\maketitle

\section{Introduction}

When the Standard Model (SM) is considered an effective low-energy theory, physics beyond the SM (BSM) can be encoded in higher-dimensional operators that supplement the SM Lagrangian
but still respect the $SU(3)\times SU(2)_L\times U(1)$ gauge symmetries. 
At dimension $5$ only a single such operator exists, the lepton-number-violating Weinberg operator~\cite{Weinberg:1979sa}, but at dimension $6$ a host of new terms become possible~\cite{Buchmuller:1985jz,Grzadkowski:2010es}. Among these are operators that involve quark currents $\bar q_f \Gamma q_i$, with possible Dirac structures 
$\Gamma\in\{\unity,\gamma_5,\gamma^\mu,\gamma^\mu\gamma_5,\sigma^{\mu\nu}\}$. For the calculation of low-energy observables the matrix elements of these operators
are often crucial input quantities, both in mesonic and baryonic systems. However, due to the absence of scalar, pseudoscalar, or tensor probes in the SM, only the matrix elements
of vector and axial-vector quark currents can be directly taken from experiment.

Scalar and pseudoscalar operators are related to vector and axial-vector ones by Ward identities~\cite{GellMann:1960np,GellMann:1964tf,Glashow:1967rx} 
\begin{align}
 \partial_\mu (\bar q_f \gamma^\mu q_i)&=i(m_f-m_i)\bar q_f q_i,\notag\\
 \partial_\mu (\bar q_f \gamma^\mu\gamma_5 q_i)&=i(m_f+m_i)\bar q_f\gamma_5 q_i,
\end{align}
known as the conservation of the vector current (CVC) and partial conservation of the axial current (PCAC), which sometimes imply relations among matrix elements, e.g.\ in the context
of nuclear $\beta$ decay~\cite{Gonzalez-Alonso:2013ura}. Similarly, the chiral symmetry of QCD provides constraints in the form of low-energy theorems, 
corrections to which can be systematically studied in chiral perturbation theory~\cite{Weinberg:1978kz,Gasser:1983yg,Gasser:1984gg},
e.g.\ the Callan--Treiman low-energy theorem for $K_{\ell 3}$ and $K_{\ell4}$ form factors~\cite{Callan:1966hu,Weinberg:1966zz,Dashen:1969bh,Bernard:2006gy,Abbas:2009dz,Colangelo:2015kha}
or the Cheng--Dashen low-energy theorem for the pion--nucleon $\sigma$-term~\cite{Cheng:1970mx,Brown:1971pn,Gasser:1990ce,Hoferichter:2015dsa}.
In contrast, for the tensor current
even in the meson sector unknown low-energy constants appear already at leading order~\cite{Cata:2007ns}, so that in general matrix elements of the quark tensor current 
are difficult to constrain from experiment, both directly and indirectly, with the only indirect connection via moments of parton distribution functions~\cite{Diehl:2003ny,Courtoy:2015haa,Kang:2015msa,Ye:2016prn,Lin:2017stx,Radici:2018iag}.
Accordingly, these matrix elements are critical input quantities for BSM searches in a variety of processes, including
kaon~\cite{Colangelo:1999kr,Antonelli:2008jg} and $\tau$ decays~\cite{Devi:2013gya,Cirigliano:2017tqn,Miranda:2018cpf}, dark matter searches~\cite{Goodman:2010ku,Hoferichter:2015ipa,Hoferichter:2016nvd,Bishara:2017pfq,Hoferichter:2018acd}, 
$\mu\to e$ conversion in nuclei~\cite{Cirigliano:2009bz,Crivellin:2014cta,Crivellin:2017rmk,Cirigliano:2017azj}, and electric dipole moments~\cite{Pospelov:2005pr,Engel:2013lsa}. 

In recent years precise calculations of the tensor charges from lattice QCD have become available both for mesons~\cite{Becirevic:2000zi,Baum:2011rm} and the nucleon~\cite{Bhattacharya:2015esa,Alexandrou:2017qyt,Yamanaka:2018uud,Gupta:2018lvp}, 
which thus determine certain form factor normalizations.
Depending on the application also the momentum dependence becomes relevant, but to constrain this behavior additional information is available from analyticity and unitarity of the form factors. In fact, 
as long as a single intermediate state dominates the unitarity relation, an exact representation can be given in terms of amplitudes that at least in principle 
are accessible in experiment as well as the normalization as determined from lattice QCD. Physically, such relations come about 
because a vector resonance can be described equivalently by a vector or an antisymmetric tensor field~\cite{Ecker:1988te,Ecker:1989yg}, 
in such a way that the same hadronic resonances appear in both form factors.
This strategy has been used recently in the context of $\tau$ decays~\cite{Cirigliano:2017tqn,Miranda:2018cpf}.

In this Letter we focus on the nucleon matrix elements. First, we consider $\pi\pi$ intermediate states and derive the corresponding unitarity relation, 
in analogy to the electromagnetic form factors~\cite{Frazer:1960zza,Frazer:1960zzb,Hohler:1976ax,Mergell:1995bf,Belushkin:2006qa,Lorenz:2014yda}. 
In combination with a narrow $\omega$ resonance this determines all $2\pi$ and $3\pi$ contributions corresponding to $J^{PC}=1^{--}$ quantum numbers. 
As first pointed out in~\cite{Gamberg:2001qc}, there are also contributions from the $1^{+-}$ channel, dominated by the $h_1(1170)$ and $b_1(1235)$ resonances for isospin $I=0$ and $I=1$, 
which mainly couple to $3\pi$ and $4\pi$ intermediate states, respectively. Combining all information from lattice-QCD tensor charges, analyticity, unitarity, and pole dominance, we provide 
a complete description of the nucleon tensor form factors below $\sqrt{|t|}\lesssim 1\GeV$.

\section{Meson form factors}

\begin{figure}[t]
 \centering
 \includegraphics[width=0.7\linewidth,clip]{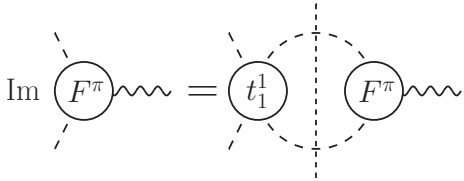}
 \caption{Elastic unitarity relation for the pion form factors $F^\pi=\{F_\pi^V,B_T^{\pi,q}\}$. The dashed lines denote pions, the wiggly lines the external current, and the short-dashed line indicates that the intermediate-state particles are to be taken on-shell.}
 \label{fig:unitarity_pipi}
\end{figure}

\begin{figure}[t]
 \centering
 \includegraphics[width=\linewidth,clip]{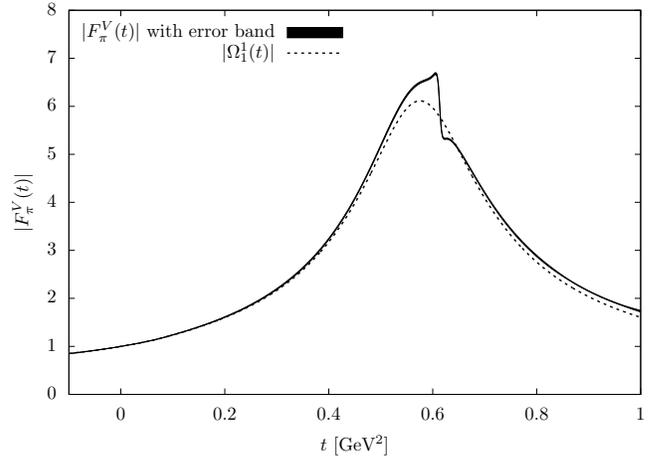}
 \caption{Comparison of the Omn\`es factor $\Omega^1_1$ to the full result for $F_\pi^V$~\cite{Colangelo:2018mtw}.}
 \label{fig:VFF_Omnes}
\end{figure}

The standard decomposition of the pion matrix element of the electromagnetic current
\beq
j^\mu_\text{em}=\sum_{q=u,d,s}Q_q\bar q \gamma^\mu q,\qquad Q=\frac{1}{3}\text{diag}\big(2,-1,-1\big), 
\eeq
reads 
\beq
\langle \pi^+(p')|j^\mu_\text{em}|\pi^+(p)\rangle=(p+p')^\mu F_\pi^V(t),\qquad t=(p'-p)^2,
\eeq
where charge conservation determines the normalization of the form factor $F_\pi^V(0)=1$. 
Elastic unitarity gives the imaginary part from $\pi\pi$ intermediate states, see Fig.~\ref{fig:unitarity_pipi},
\beq
\label{unitarity_pipi}
\Im F_\pi^V(t)=\sigma_\pi(t) \big(t^1_1(t)\big)^*F_\pi^V(t),
\eeq
where $\sigma_\pi(t)=\sqrt{1-4\mpi^2/t}$ and $t^1_1(t)$ is the $P$-wave $\pi\pi$ partial-wave amplitude related to the phase shift $\delta^1_1(t)$ according to
\beq
t^1_1(t)=\frac{1}{\sigma_\pi(t)} e^{i\delta^1_1(t)}\sin\delta^1_1(t).
\eeq
Equation~\eqref{unitarity_pipi} is a manifestation of Watson's final-state theorem~\cite{Watson:1954uc} that equates the phase of the form factor with $\delta^1_1(t)$, leading to a representation in terms 
of the Omn\`es function~\cite{Omnes:1958hv}
\beq
F_\pi^V(t)=\Omega^1_1(t),\qquad \Omega^1_1(t)=\exp\bigg\{\frac{t}{\pi}\int_{4\mpi^2}^\infty\diff t'\frac{\delta^1_1(t')}{t'(t'-t)}\bigg\}.
\eeq
In practice, this representation does not fully capture all properties of $F_\pi^V(t)$: corrections arise from inelastic states such as $3\pi$ in the vicinity of the $\omega(782)$ resonance (isospin-violating) 
as well as $4\pi$ above $1\GeV$ (isospin-conserving). These effects can be accounted for by suitable extensions of the Omn\`es representation, see~\cite{DeTroconiz:2001rip,Leutwyler:2002hm,Colangelo:2003yw,deTroconiz:2004yzs,Hanhart:2012wi,Ananthanarayan:2013zua,Ananthanarayan:2016mns,Hoferichter:2016duk,Hanhart:2016pcd,Colangelo:2018mtw},
but due to the dominance of the $\rho(770)$ the Omn\`es factor provides the bulk of the contribution, see Fig.~\ref{fig:VFF_Omnes}. In particular, the low-energy parameters are well reproduced, 
e.g.\ the radius
\beq
\label{rV}
\langle (r_\pi^V)^2\rangle=6\frac{\diff F_\pi^V(t)}{\diff t}\bigg|_{t=0}=0.419\fm^2
\eeq
differs from the full result~\cite{Colangelo:2018mtw}  
\beq
\label{rV_full}
\langle (r_\pi^V)^2\rangle=0.429(4)\fm^2
\eeq
by about $2\%$, where the increase with respect to~\eqref{rV} comes from inelastic effects.

For the decomposition of the pion matrix element of the tensor current we take~\cite{Baum:2011rm,Diehl:2005rn}
\beq
\langle \pi^+(p')|\bar q \sigma^{\mu\nu} q|\pi^+(p)\rangle=\frac{i}{\mpi}\big(p'^\mu p^\nu-p^\mu p'^\nu\big) B_T^{\pi,q}(t).
\eeq
The crucial observation is that elastic unitarity produces exactly the same result as in~\eqref{unitarity_pipi}
\beq
\Im B_T^{\pi,q}(t)=\sigma_\pi(t) \big(t^1_1(t)\big)^*B_T^{\pi,q}(t),
\eeq
so that up to inelastic corrections we may write
\beq
\label{BT_pipi}
B_T^{\pi,q}(t)=B_T^{\pi,q}(0)F_\pi^V(t).
\eeq
In principle, analyticity and unitarity alone would allow for an arbitrary polynomial multiplying the Omn\`es factor,
but in the same way as for the vector form factor constraints from perturbative QCD~\cite{Lepage:1980fj}   
should lead to an asymptotic $1/t$ behavior. 
With the normalization determined from lattice QCD~\cite{Baum:2011rm}\footnote{All tensor matrix elements are quoted at an $\overline{\text{MS}}$ scale $\mu=2\GeV$.}
\beq
\label{BT}
B_T^{\pi,u}(0)=-B_T^{\pi,d}(0)=0.195(10),
\eeq
this essentially determines the tensor form factor throughout $\sqrt{t}\lesssim 1\GeV$ except for close to the $\omega$ resonance. In particular, we obtain for the radius
\beq
\langle (r_T^{\pi,u})^2\rangle=\frac{6}{B_T^{\pi,u}(0)}\frac{\diff B_T^{\pi,u}(t)}{\diff t}\bigg|_{t=0}=0.43(1)\fm^2,
\eeq
where we have adopted the central value from~\eqref{rV_full}, assuming that inelastic effects modify $\langle (r_T^{\pi,u})^2\rangle$ in a similar way
as $\langle (r_\pi^V)^2\rangle$, and the error is motivated by the corresponding difference between~\eqref{rV_full} and ~\eqref{rV}. 
Phenomenologically, these relations work because of the dominance of the $\rho(770)$.

A similar argument applies to the flavor-non-diagonal current $\bar s \sigma^{\mu\nu} u$, whose $\pi K$ form factor
\beq
\langle \overline{K^0}(p')|\bar s \sigma^{\mu\nu} u|\pi^+(p)\rangle=\frac{i}{\mk}\big(p'^\mu p^\nu-p^\mu p'^\nu\big) B_T^{\pi K}(t)
\eeq
is related by elastic unitarity to the $K_{\ell 3}$ form factor $f_+(t)$~\cite{Antonelli:2008jg}
\beq
B_T^{\pi K}(t)=B_T^{\pi K}(0)\frac{f_+(t)}{f_+(0)},
\eeq
and inelastic corrections are suppressed with respect to the $K^*(892)$ resonance. With normalization $B_T^{\pi K}(0)=0.686(25)$~\cite{Baum:2011rm}
and $\diff f_+(t)/\diff t=f_+(0)\lambda_+'/\mpi^2$ as measured in $K_{\ell 3}$ experiments, $\lambda_+'=25.2(9)\times 10^{-3}$~\cite{Antonelli:2008jg}, this gives
\beq
\langle (r_T^{\pi K})^2\rangle=\frac{6}{B_T^{\pi K}(0)}\frac{\diff B_T^{\pi K}(t)}{\diff t}\bigg|_{t=0}=0.30(2)\fm^2,
\eeq
where the uncertainty derives from the comparison of the physical $\lambda_+'$ with the derivative of the $\pi K$ Omn\`es factor
as an estimate of the impact of inelastic effects.

Finally, we note that the lattice normalizations determine some of the low-energy constants $\Lambda_i$ from~\cite{Cata:2007ns}. The $\pi\pi$ result~\eqref{BT} implies
$\Lambda_2=\Fpi^2B_T^{\pi,u}(0)/(2\mpi)=6.0(3)\MeV$~\cite{Dekens:2018pbu,Tanabashi:2018oca}, in good agreement with the $SU(3)$-related version
$\Lambda_2=\Fpi^2B_T^{\pi K}(0)/(2\mk)=5.9(2)\MeV$.
Further, by assuming $\rho$ pole dominance, we obtain the relation (see also~\cite{Dekens:2018pbu})
\beq
\Lambda_1=-\frac{M_\rho^2 B_T^{\pi,u}(0)}{4g_{\rho\pi\pi}g_{\rho\gamma}\mpi}=-7.1(7)\MeV
\eeq
in terms of the $\rho\pi\pi$ and $\rho\gamma$ coupling constants~\cite{Colangelo:2001df,GarciaMartin:2011jx,Hoferichter:2017ftn},
and we have attached a $10\%$ uncertainty as suggested by the analog estimate for the charge radius $\langle (r_{\pi}^{V,\rho})^2\rangle = 6/M_\rho^2=0.39\fm^2$.

\section{Nucleon form factors}

\begin{figure}[t]
 \centering
 \includegraphics[width=0.7\linewidth,clip]{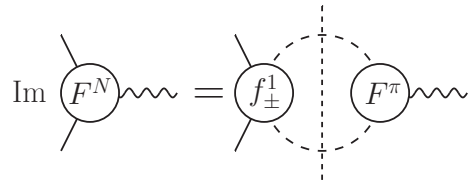}
 \caption{Elastic unitarity relation for the isovector nucleon form factors $F^N=\{F_i,F_{i,T}^q\}$ and accordingly $F^\pi=\{F_\pi^V,B_T^{\pi,q}\}$. The solid lines denote nucleons, otherwise notation as in Fig.~\ref{fig:unitarity_pipi}.}
 \label{fig:unitarity_piN}
\end{figure}

The nucleon transition matrix element of the electromagnetic current operator has
the standard form
\beq
\langle N(p')|j^\mu_\text{em}|N(p)\rangle=\bar u(p')\Big[\gamma^\mu F_1(t) + \frac{i \sigma^{\mu\nu}q_\nu}{2\mN} F_2(t)\Big]u(p),
\eeq
where $q=p'-p$. For the isovector combinations $F_i^v=(F_i^p-F_i^n)/2$, $\pi\pi$ intermediate states again dominate the unitarity relation, in such a way that the strategy from the mesonic system generalizes in
a straightforward way to the nucleon case, see Fig.~\ref{fig:unitarity_piN}. The main difference, however, concerns the fact that on the right-hand side of the unitarity relation both the scattering amplitude, $\pi\pi\to N\bar N$ in this case, and the pion form factor enter as input. Following the notation from~\cite{Hoferichter:2015hva,Hoferichter:2016duk}, we have~\cite{Federbush:1958zz,Frazer:1960zza,Frazer:1960zzb}
\begin{align}
\label{ImFi_em}
 \Im F_1^v(t)&=\frac{q_t^3}{2\sqrt{t}}\big(F_\pi^V(t)\big)^*\bigg(\frac{2\mN}{p_t^2}\Gamma^1(t)+\sqrt{2} f_-^1(t)\bigg),\notag\\
 \Im F_2^v(t)&=-\frac{\mN q_t^3}{p_t^2\sqrt{t}}\big(F_\pi^V(t)\big)^*\Gamma^1(t),
\end{align}
where $f^1_\pm(t)$ are the $P$-wave partial waves for $\pi\pi\to N\bar N$, 
\beq
\Gamma^1(t)=\frac{\mN}{\sqrt{2}} f_-^1(t)-f_+^1(t),
\eeq
and $p_t=\sqrt{t/4-\mN^2}$, $q_t=\sqrt{t/4-\mpi^2}$.

The matrix elements of the tensor current decompose according to~\cite{Weinberg:1958ut,Adler:1975he}
\begin{align}
 &\langle N(p')|\bar q \sigma^{\mu\nu} q|N(p)\rangle\notag\\
 &=\bar u(p')\Big[\sigma^{\mu\nu} F_{1,T}^q(t) + \frac{i}{\mN}\big(\gamma^\mu q^\nu-\gamma^\nu q^\mu\big)F^q_{2,T}(t)\notag\\
 &\qquad+\frac{i}{\mN^2}\big(P^\mu q^\nu-P^\nu q^\mu\big)F^q_{3,T}(t)\Big]u(p)\notag\\
 &=\bar u(p')\Big[\sigma^{\mu\nu} F_{1,T}^q(t) + \frac{i}{\mN}\big(\gamma^\mu q^\nu-\gamma^\nu q^\mu\big)\hat F^q_{2,T}(t)\notag\\
 &\qquad+\frac{1}{\mN^2}\big(\sigma^{\mu\alpha} q^\nu-\sigma^{\nu\alpha} q^\mu\big)q_\alpha F^q_{3,T}(t)\Big]u(p),
\end{align}
where $P=p+p'$ and $\hat F_{2,T}^q(t)=F_{2,T}^q(t)+2F_{3,T}^q(t)$. As expected, the evaluation of the unitarity relation of the tensor form factors from $\pi\pi$ intermediate states produces a result that only depends on the $P$-wave amplitudes. We find
\begin{align}
\label{FT_spectral}
\Im F_{1,T}^{q,v}(t)&=0,\notag\\
 \Im F_{2,T}^{q,v}(t)&=-\frac{\mN q_t^3}{2\mpi\sqrt{2t}}\big(B_T^{\pi,q}(t)\big)^*f_-^1(t),\notag\\
 \Im F_{3,T}^{q,v}(t)&=-\frac{\mN^2 q_t^3}{4\mpi p_t^2\sqrt{t}}\big(B_T^{\pi,q}(t)\big)^*\Gamma^1(t).
\end{align}
Using~\eqref{ImFi_em} and~\eqref{BT_pipi}, this leads to the identification 
\begin{align}
\label{FT_v}
F_{1,T}^{q,v}(t)&=0,\notag\\
F_{2,T}^{q,v}(t)&=-\frac{\mN}{2\mpi}B_T^{\pi,q}(0)\big(F_1^v(t)+F_2^v(t)\big),\notag\\
F_{3,T}^{q,v}(t)&=\frac{\mN}{4\mpi}B_T^{\pi,q}(0)F_2^v(t),
\end{align}
for the $I=1$, $J^{PC}=1^{--}$ contribution, valid up to inelastic corrections. In particular, vector intermediate states do not contribute to $F_{1,T}^{q,v}$. 
As a first check on these relations we consider the tensor anomalous magnetic moments $\kappa_T^q=-2\hat F_{2,T}^q(0)$
and compare to lattice QCD~\cite{Gockeler:2006zu}
\beq
\kappa_T^{u,p}\approx 3.0,\qquad \kappa_T^{d,p}\approx 1.9,
\eeq
in reasonable agreement with~\eqref{FT_v}
\beq
\label{kappa_v}
1.1\approx \kappa_T^{u,p}-\kappa_T^{u,n}=\frac{\mN}{\mpi} B_T^{\pi,u}(0)=1.31(7),
\eeq
where we have assumed isospin symmetry $\kappa_T^{u,n}=\kappa_T^{d,p}$ etc.
In fact, the identification~\eqref{FT_v} is precisely what one would have obtained assuming a narrow resonance to describe the $\rho$: using the Lagrangian from~\cite{Kubis:2006cy} we find
\beq
F_1^{v,\rho}(t)=\frac{1}{2} D_\rho(t),\qquad F_2^{v,\rho}(t)=\frac{\kappa_\rho}{2} D_\rho(t),
\eeq
with $D_\rho(t)=M_\rho^2/(M_\rho^2-t)$, where $g_\rho=g_{\rho\gamma}$ has been assumed to ensure the correct normalization of $F_1^v$
and $\kappa_\rho$ parameterizes the magnetic-moment coupling of the $\rho$ in the conventions of~\cite{Kubis:2006cy}.
Moreover, $\rho$ pole dominance gives for the ratio of tensor and vector coupling constants
\beq
\frac{F_\rho^T}{F_\rho}=\frac{B_T^{\pi,u}(0)}{2}\frac{M_\rho}{M_\pi}=0.54(3),
\eeq
in reasonable agreement with the most recent result from lattice QCD $F_\rho^T/F_\rho=0.629(8)$~\cite{Braun:2016wnx} and not far from the expectation $1/\sqrt{2}$ from large $N_c$~\cite{Cata:2008zc}.
Accordingly, the calculation of the nucleon tensor form factors then reproduces~\eqref{FT_v}, in such a way that the dispersive derivation in terms of the spectral functions should be considered a more rigorous justification that does not rely on a narrow-resonance assumption, only on the dominance of elastic intermediate states. 

Beyond the isovector combination, these arguments for~\eqref{FT_v} suggest to estimate the isoscalar $1^{--}$ contributions in a similar way based on a narrow $\omega$ as a description of the $3\pi$ channel
\beq
F_1^{s,\omega}(t)=\frac{1}{2} D_\omega(t),\qquad F_2^{s,\omega}(t)=\frac{\kappa_\omega}{2} D_\omega(t).
\eeq
Lattice results for $F_\phi^T/F_\phi$ indicate a deviation from the $\rho$ around $10\%$~\cite{Braun:2003jg,Allton:2008pn,Jansen:2009hr},
which in turn implies that $F_\omega^T/F_\omega$ should be very close as well, given that the $SU(3)$ corrections are absent while the small Okubo--Zweig--Iizuka rule violations only increase
by a factor $2$~\cite{Cata:2009dq}. In this way, we obtain an isovector analog of~\eqref{FT_v} up to an overall factor of $g_{\omega\gamma}/g_{\rho\gamma}=3$. 
Indeed, the comparison to the isoscalar combination of tensor anomalous magnetic moments
\beq
\label{kappa_s}
4.9\approx \kappa_T^{u,p}+\kappa_T^{u,n}=3\frac{\mN}{\mpi} B_T^{\pi,u}(0)=3.94(20)
\eeq
works at a similar level as~\eqref{kappa_v}, and both are in reasonable agreement given 
the exploratory character of the lattice results~\cite{Gockeler:2006zu} and the fact that the dispersive derivation relies on an unsubtracted dispersion relation. 
Note that even for the channels for which no rigorous spectral functions as in~\eqref{FT_spectral} are known, 
an ansatz for the full momentum dependence with good analytic properties
can be made by replacing the narrow-width propagators by a dispersion relation
\beq
\frac{1}{\pi}\int_{t_\text{thr}}^\infty\diff t'\frac{\Im d_R(t')}{t'-t},\qquad d_R(t)=\frac{1}{M_R^2-t-i M_R \Gamma_R},
\eeq
or variants thereof, e.g.\ with an energy-dependent width~\cite{Lomon:2012pn,Moussallam:2013una,Hoferichter:2014vra,Hoferichter:2018kwz}.

Next, we turn to the contributions from the $C$-odd axials $h_1(1170)$ and $b_1(1235)$~\cite{Gamberg:2001qc,Ecker:2007us}. Modifying the Lagrangian from~\cite{Kubis:2006cy} by Levi-Civita  tensors to account for the different parity, we find that only one structure produces a resonant contribution
\begin{align}
 F_{1,T}^{q,a}(t)&=F_{1,T}^{q}(0) D_R(t),\qquad \hat F_{2,T}^{q,a}(t)=0,\notag\\
 F_{3,T}^{q,a}(t)&=-F_{1,T}^q(0)\frac{\mN^2}{M_R^2} D_R(t),
\end{align}
where $R=h_1(1170)$ and $R=b_1(1235)$ for the isoscalar and isovector combinations, respectively, and the superscript $a$ indicates the new axial-vector contribution  
(parity forbids such a coupling for the pseudoscalar mesons). 
The tensor charges~\cite{Gupta:2018lvp} determine the normalizations according to
\beq
F_{1,T}^{u,p}(0)=0.784(28),\qquad F_{1,T}^{d,p}(0)=-0.204(11).
\eeq
Taking together vector and axial-vector contributions, we find the representation
\begin{align}
\label{FT_final}
 F_{1,T}^{q,p}(t)&=\pm \frac{1}{2}\big(F_{1,T}^{u,p}(0)-F_{1,T}^{d,p}(0)\big)D_{b_1}(t)\\
 &+\frac{1}{2}\big(F_{1,T}^{u,p}(0)+F_{1,T}^{d,p}(0)\big)D_{h_1}(t),\notag\\
 F_{2,T}^{u,p}(t)&=-\frac{\mN}{2\mpi}B_T^{\pi,u}(0)\big(2G_M^p(t)+G_M^n(t)\big)+2\tilde F_{1,T}^{u,p}(t),\notag\\
 F_{2,T}^{d,p}(t)&=-\frac{\mN}{2\mpi}B_T^{\pi,u}(0)\big(G_M^p(t)+2G_M^n(t)\big)+2\tilde F_{1,T}^{d,p}(t),\notag\\
 F_{3,T}^{u,p}(t)&=\frac{\mN}{4\mpi}B_T^{\pi,u}(0)\big(2F_2^p(t)+F_2^n(t)\big)
 -\tilde F_{1,T}^{u,p}(t),\notag\\
 F_{3,T}^{d,p}(t)&=\frac{\mN}{4\mpi}B_T^{\pi,u}(0)\big(F_2^p(t)+2F_2^n(t)\big)
 -\tilde F_{1,T}^{d,p}(t),\notag
\end{align}
where $\pm$ corresponds to $q=u,d$, $G_M=F_1+F_2$, $\tilde F_{1,T}^{q}$ is $F_{1,T}^{q,a}(t)$ multiplied by $\mN^2/M_R^2$, 
and the neutron form factors follow from isospin symmetry. Our final results for normalizations and slopes according to~\eqref{FT_final} are collected in Table~\ref{tab:numbers}.
The uncertainties are estimated as follows:
first, the part of the normalizations of $F_{2,T}^q$ and $F_{3,T}^q$ derived from the electromagnetic form factors is assigned a $40\%$ uncertainty, 
corresponding to the sum-rule violations observed in~\cite{Hoferichter:2016duk}. Similarly, the isovector sum rules for the electromagnetic radii 
suggest an accuracy of the derivatives at a level of $10\%$. In both cases, the isoscalar extension to the $\omega$ (and $\phi$) should hold at a similar level.
This expectation follows from the fact that the reason for the slow convergence of the isovector sum rules as well as
the departure from simple pole dominance for the $\rho$, e.g.\ compared to 
the meson form factors discussed above, traces back to the singularity structure of the nucleon Born terms in $\pi\pi\to N\bar N$,
but this threshold enhancement is compensated by phase space for intermediate states with higher multiplicity such as $3\pi$~\cite{Bernard:1996cc}.
In the same way, since the $1^{+-}$ resonances mainly couple to the $3\pi$ and $4\pi$ channels, pole dominance should again work reasonably well, as suggested 
by the meson examples we attach a $20\%$ uncertainty. Indeed, the corresponding slopes agree well with 
$\dot F_{1,T}^{u,p}=0.57(3)\GeV^{-2}$ and $\dot F_{1,T}^{d,p}=-0.14(2)\GeV^{-2}$ from lattice QCD~\cite{Gockeler:2005cj} (note that these errors are incomplete e.g.\ due to disconnected diagrams).
In all cases, the uncertainties associated with sum-rule convergence and pole dominance by far outweigh the uncertainties in the nucleon form factors. 
For definiteness, we take magnetic moments and $\langle r_E^2\rangle^n=-0.12\fm^2$ from~\cite{Tanabashi:2018oca} (the latter mainly based on~\cite{Koester:1995nx,Kopecky:1997rw})
as well as $r_E^p=0.84\fm$~\cite{Pohl:2010zza,Antognini:1900ns}, $r_M^p=0.87\fm$, $r_M^n=0.89\fm$~\cite{Epstein:2014zua}.

\begin{table}[t]
\renewcommand{\arraystretch}{1.3}
\centering
\begin{tabular}{lrr}\toprule
Form factor & Normalization  & Radius $[\text{fm}^2]$\\\colrule
$B_T^{\pi,u}=-B_T^{\pi,d}$ & $0.195(10)$ & $0.43(1)$\\
$B_T^{\pi K}$ & $0.686(25)$ & $0.30(2)$\\\colrule
Form factor & Normalization  & Slope $[\text{GeV}^{-2}]$\\\colrule
$F_{1,T}^{u,p}$ & $0.784(28)$ & $0.54(11)$\\
$F_{1,T}^{d,p}$ & $-0.204(11)$ & $-0.11(2)$\\
$F_{2,T}^{u,p}$ & $-1.5(1.0)$ & $-7.0(8)$\\
$F_{2,T}^{d,p}$ & $0.5(3)$ & $2.5(3)$\\
$F_{3,T}^{u,p}$ & $0.1(2)$ & $1.8(2)$\\
$F_{3,T}^{d,p}$ & $-0.6(3)$ & $-2.1(2)$\\\colrule
$F_{1,T}^{s,N}$ & $-0.0027(16)$ & $-0.0014(9)$\\
$F_{2,T}^{s,N}$ & $0.009(5)$ & $0.041(26)$\\
$F_{3,T}^{s,N}$ & $-0.004(3)$ & $-0.015(13)$\\
\botrule
\end{tabular}
\caption{Summary of normalizations and radii/slopes for the meson and nucleon tensor form factors.}
\label{tab:numbers}
\end{table}

Finally, we remark that assuming $SU(3)$ symmetry for the tensor coupling the strangeness form factors can be estimated in close analogy to~\eqref{FT_final}
\begin{align}
\label{FT_s}
 F_{1,T}^{s,N}(t)&=F_{1,T}^{s,N}(0)D_{h_1^s}(t),\notag\\
 F_{2,T}^{s,N}(t)&=-\frac{\mN}{2\mpi}B_T^{\pi,u}(0)G_M^{s,N}(t)+2\tilde F_{1,T}^{s,N}(t),\notag\\
 F_{3,T}^{s,N}(t)&=\frac{\mN}{4\mpi}B_T^{\pi,u}(0)F_2^{s,N}(t)-\tilde F_{1,T}^{s,N}(t),
\end{align}
where $h_1^s=h_1(1380)$. Recent studies in lattice QCD show that the vector~\cite{Alexandrou:2019olr} and tensor~\cite{Gupta:2018lvp} strangeness content of the nucleon is tiny, leading to the estimates
in Table~\ref{tab:numbers}. In fact, \eqref{FT_s} even predicts that $\kappa_T^{s,N}\equiv 0$. For strangeness, the uncertainties in the vector matrix elements need to be included, 
we use $\mu^s=-0.017(4)$, $\langle r_{E,s}^2\rangle^N=-0.0048(6)\fm^2$, $\langle r_{M,s}^2\rangle^N=-0.015(9)\fm^2$~\cite{Alexandrou:2019olr}.
The matrix elements of the heavy quarks $c,b,t$ can be addressed using the heavy-quark expansion~\cite{Polyakov:2015foa}.

In conclusion, our main results in~\eqref{FT_final} and Table~\ref{tab:numbers} summarize the present status of meson and nucleon form factors of the antisymmetric tensor current, using all available information from lattice QCD, analyticity, unitarity, and pole dominance. While in general the momentum dependence of the form factors can be reconstructed quite accurately, for the normalizations input from lattice QCD is key, and the dominant uncertainties in the normalization of the induced nucleon form factors 
precisely reflect the fact 
that the sum rules for the normalization exhibit rather slow convergence. However, the combination of all available information 
allows us to provide reliable results both for the normalizations and momentum dependence, 
including form factors that previously could only be estimated using hadronic models~\cite{Pasquini:2005dk,Ledwig:2010zq,Erkol:2011iw,Ledwig:2011qw}.
Applications are immediate for the search for dark matter in direct detection experiments and lepton flavor violation in $\mu\to e$ conversion in nuclei.
In addition, the method developed here could be extended to hyperon tensor form factors, and thus improve the matrix elements required 
in the search for non-standard hyperon decays~\cite{Chang:2014iba}.

\begin{acknowledgments} 
We thank Gilberto Colangelo, Aurore Courtoy, and Ulf-G.\ Mei\ss ner for comments on the manuscript.
Financial support by the DOE (Grant Nos.\ DE-FG02-00ER41132 and DE-SC0009919), 
the DFG and NSFC through funds provided to the Sino--German CRC 110
``Symmetries and the Emergence of Structure in QCD'' (DFG
Grant  No.\  TRR110  and  NSFC  Grant  No.\  11621131001),
and the Swiss National Science Foundation
(Project Nos.\ PZ00P2\_174228 and P300P2\_167751) is gratefully acknowledged.
\end{acknowledgments}

\end{document}